\documentclass[doublecol]{epl2}

\usepackage{graphicx,bbm,setspace,amssymb,amsfonts,amsmath,mathtools,mathrsfs,stmaryrd,amsthm,lingmacros,enumitem,algorithm,nicefrac,bm,caption,subcaption,etoolbox,comment,url,soul}

\usepackage{appendix}

\usepackage{dcolumn}

\DeclareMathOperator*{\argmin}{argmin}

\DeclareMathOperator{\E}{\mathbb{E}}

\newtheorem{theorem}{Theorem}
\newtheorem{thm-defn}[theorem]{Theorem/Definition}

\newtheorem{prop}[theorem]{Proposition}

\theoremstyle{definition}

\theoremstyle{remark}

\title{Understanding spatial propagation using metric geometry with application to the spread of COVID-19 in the United States}
\shorttitle{Understanding spatial propagation using metric geometry} 

\author{N. James\inst{1} \and M. Menzies \inst{2} \and H. Bondell\inst{1}}
\shortauthor{N. James \etal}

\institute{                    
  \inst{1} School of Mathematics and Statistics, University of Melbourne, Victoria 3010, Australia\\
  \inst{2} Yau Mathematical Sciences Center, Tsinghua University, Beijing 100084, China
}

\abstract{
This paper introduces a novel approach to spatio-temporal data analysis using metric geometry to study the propagation of COVID-19 across the United States. Using a geodesic Wasserstein metric, we analyse discrepancies between the density functions of new case counts on any given day, incorporating the geographic spread of cases. First, we apply this to identify the periods during which the changes in the geographic distribution of COVID-19 were most profound. The greatest shift occurred between May and June of 2020, when COVID-19 shifted from mostly dominating the Northeastern states to a wider distribution across the country. We support our findings with a new measure of the extent of geodesic variance of a distribution, demonstrating that the geographic imprint of COVID-19 was most concentrated in May 2020. Next, we investigate whether the epidemic exhibited meaningful patterns of spatial reversion, where similar geographic distributions return later. We identify broad similarity between the spread of COVID-19 across the US between the second and third waves, and to a lesser extent, the reemergence of the first wave's Northeastern dominance closer to the present day. This methodology could provide new insights for analysts to monitor the dynamical spread of epidemics and enable regional policymakers to protect their localities. More broadly, the framework we introduce could be applied to a variety of problems evolving over space and time.
}

\begin{document}

\maketitle


\section{Introduction}

COVID-19 remains an ongoing threat to the public health of the United States (US). While an aggressive vaccination program has reduced cases and deaths considerably from their highs in January 2021 \cite{vaccinereducedeaths}, the vaccination rate is highly non-uniform across the country \cite{worldindata2020_vax,cnnvaxnonuniform}. Communities with low vaccination rates remain a risk, to themselves, as well as to the rest of the nation, both regarding the spread of the virus to vulnerable people elsewhere and the potential for further virus mutation. Throughout the pandemic, the imposition of restrictions on community activity and businesses, including mask mandates and lockdowns, has mostly fallen to state and local governments. Thus, it remains of considerable interest to local lawmakers to track the ongoing spatial propagation of COVID-19 across the US. This may reveal trends in the spread of the virus, allow for forewarning before local outbreaks, identify locations that performed unusually well or poorly in containment measures, and provide opportunities to learn from other localities.

There are numerous existing analyses of the propagation of COVID-19 across the US on a state-by-state or finer basis. Previous research has used a variety of techniques, including state-by-state time series analysis \cite{james2020covidusa,jamescovideu}, SEIR models \cite{Zhou2020_covidUS}, regression models and feature selection \cite{Gross2020,Maiti2021}, Markov chain Monte Carlo models \cite{Paul2020}, and other Monte Carlo simulations \cite{Wang2020_spatioUS}. Similar studies have been carried out in other countries, such as Brazil \cite{Castro2021_Brazil,Ribeiro2020,daSilva2021}, China \cite{Feng2020_spatioChina} and the United Kingdom \cite{Zheng2021}, or comparing numerous countries simultaneously  \cite{james2021_mobility,James2021_virulence,Manchein2020,Jamesfincovid}. Most of these studies are qualitatively descriptive in their summary of the spread of COVID-19 across the geography of a country. 
Prior physically-inspired research on COVID-19 has focused on fluids and aerosols \cite{Ng2021_fluids,McRae2021,Verma2020,Yang2020_fluids,Abkarian2020}, characteristics of optimal antigens \cite{Ganti2021},  contagion and percolation models \cite{deArruda2020,Luo2020} and network models \cite{deArruda2020_network, Moore2020, Castellano2020}. Methods from statistical physics have been used widely to study both COVID-19 and prior epidemics \cite{Ganti2021,Silva2019,deArruda2017,Castellano2020}.

\section{Geodesic Wasserstein distance}

This paper introduces a new mathematical method for studying the spatial propagation of an epidemic over time. Our approach applies metric geometry to perform an in-depth analysis of the time-varying spatial distribution of COVID-19 cases across US states and the District of Columbia (DC). Let $(X,d)$ be any metric space, $\mu,\nu$ two probability measures on $X$, and $q \geq 1$. The Wasserstein metric between $\mu, \nu$ is defined as
\begin{align}
\label{eq:Wasserstein}
    W^q (\mu,\nu) = \inf_{\gamma} \bigg( \int_{X \times X} d(x,y)^q d\gamma  \bigg)^{\frac{1}{q}},
\end{align}
where the infimum is taken over all probability measures $\gamma$ on $X \times X$ with marginal distributions $\mu$ and $\nu$. Henceforth, let $q=1$. By the Kantorovich-Rubinstein formula \cite{Kantorovich}, there is an alternative formulation when X is compact (for example, finite):
\begin{align}
\label{eq:Wassersteinalt}
    W^1 (\mu,\nu) = \sup_{F} \left| \int_{X} F d\mu - \int_{X} F d\nu  \right|,
\end{align}
where the supremum is taken over all $1$-Lipschitz functions $F: X \to \mathbb{R}$.

Throughout this paper, let $X$ be the set of the 50 US states and DC, ordered alphabetically and indexed $i=1,...,51$. As this is a finite set, measures $\mu,\nu$ may be reinterpreted as probability vectors $f,g \in \mathbb{R}^{51}$ such that $f_i\geq 0$ and $\sum_{i=1}^{51} f_i=1$. Such probability vectors are our central object of study, as they define the spatial distribution over the US. The simplest metric we can equip $X$ with is the discrete metric, where $d(x,y)=1$ if $x\neq y$, and 0 otherwise. In this case, the formulation (\ref{eq:Wassersteinalt}) of Wasserstein metric is optimised with the choice of function
\begin{align}
\label{eq:bestF}
F(x) = \begin{cases}
1, & f(x) \geq g(x), \\
0, & f(x)<g(x), 
\end{cases}
\end{align}
and thus greatly simplifies to a relatively simple metric
\begin{align}
\label{eq:discreteWass}
     W_{disc}(f,g)=\frac12 ||f - g||_1.
\end{align}
A proof of this statement is included in the appendix (Proposition 1). We refer to this as the \emph{discrete Wasserstein distance} - it will not be our primary object of study but will be utilised later to demonstrate the robustness of our results.

Our primary methodological contribution is obtained by setting the metric $d$ on $X$ to be the real-world geodesic distances between (the centroids of) the 50 US states and DC, measured in meters. We term the associated Wasserstein distance the \emph{geodesic Wasserstein distance} and notate it as $W_{G}(f,g)$ for two probability vectors $f$ and $g$. Wasserstein geodesics have been utilised in optimal transport problems \cite{Panaretos2019,Lavenant2019}, but are novel in epidemic research, which frequently neglects the spatial aspect of a virus' spread.

Consider the illustrative case where $f$ and $g$ are Dirac delta functions supported on single elements $x$ and $y$, respectively. By the formulation (\ref{eq:Wassersteinalt}), it follows that $W^1(f,g)=d(x,y)$. For the discrete Wasserstein distance, this means that $W_{disc}(f,g)=1$ if $x\neq y$, whereas $W_{G}(f,g)$ is the physical distance between (the centroids of the) states indexed by $x$ and $y$. Thus, the geodesic Wasserstein distance allows us to factor in the physical spread of COVID-19 across the US. That is, the discrete Wasserstein distance (or $L^1$ norm) between probability vectors does not take into account distance between states; the geodesic Wasserstein awards a greater distance between two probability vectors if the proportion of cases has shifted further away geographically. Intuitively, the geodesic Wasserstein metric $W_{G}(f,g)$ is the cost or work (in the sense of physics) to transform the distribution $f$ into $g$, taking real-world distances into account. For example, a shift in distribution from the Northeast to the Appalachian states would be granted a smaller discrepancy difference than a shift from the Northeast to Pacific states.

We use this new metric for two aims. First, we wish to determine the periods with the most rapid change in spatial propagation of COVID-19 across the US. Second, we investigate whether there is any reversion behaviour in this spatial propagation. While existing studies have typically analysed the behaviour of the different waves of the pandemic on a state-by-state basis, our approach will identify distinct times where a similar geographic distribution of COVID-19 returned across the US. In doing so, it can identify similarities in successive waves of the pandemic.

\section{Temporal changes in spatial propagation}

\begin{figure}
    \centering
    \includegraphics[width=0.49\textwidth]{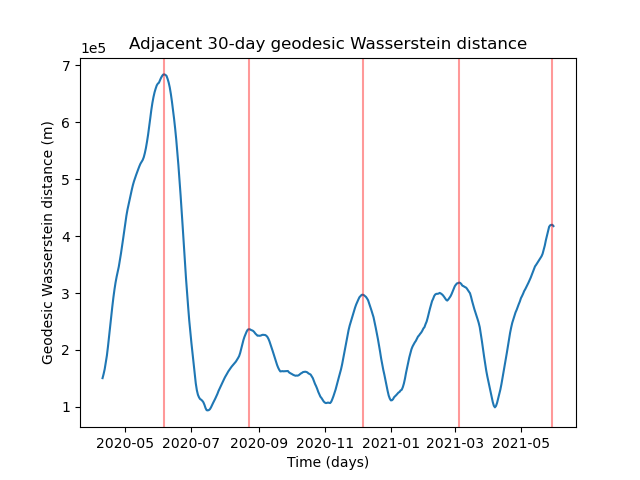}
    \caption{Time-varying geodesic Wasserstein distance (in meters) between adjacent aggregated 30-day distributions of new cases. Large spikes in this distance indicate meaningful shifts in the spatial distribution of COVID-19 over a one-two month transition. The large first peak corresponds to June 6, 2020, signifying that the prior and following 30-day periods have substantially different spatial distributions. The prior 30 days (May 7 to June 5) are characterised by the dominance of Northeastern states, while the following 30 days (June 6 to July 5) are characterised by the ascent of case numbers elsewhere, including California, Texas, Florida and Arizona.}
    \label{fig:GW_CPD}
\end{figure}

Our data spans March 12, 2020 to June 30, 2021  across $n=51$ regions (50 states and DC), a period of $T=476$ days. We begin here to avoid periods of sparse reporting early in the pandemic. Our analysis includes Alaska and Hawaii - due to their small contribution to total US case counts, their exclusion has a minimal impact on results. In order to reduce irregularities in daily counts, such as lower reporting of tests on weekends, we first apply a simple 7-day smoothing operator to the counts. This yields a multivariate time series of smoothed new cases $x_i(t), i=1,...,n, t=1,..,T$.  For the first experiment, we consider grouped probability vectors of 30-day periods. Specifically, let $f^{[a:b]}$ be the probability vector of (smoothed) new cases in each state, observed across an interval $a\leq t \leq b$, divided by the total number of US cases across this period:
\begin{align}
f^{[a:b]}_i = \dfrac{\sum_{t=a}^b x_i(t)}{\sum_{t=a}^b \sum_{j=1}^{n} x_j(t)}, i={1},...,n. 
\end{align}
When $a=b$, we simply notate this as $f_i(t)$. In fig. \ref{fig:GW_CPD}, we display the distance function 
\begin{align}
\label{eq:distancefn1}
   t \mapsto W_G(f^{[t-30:t-1]}, f^{[t:t+29]}), \hspace{1em} 31 \leq t \leq T - 29,
\end{align}
 between adjacent 30-day periods. For example, $t=51$ corresponds to May 1, and the associated function value is the geodesic Wasserstein distance between the probability vector of cases across April 1-30 and May 1-30. Within this figure, we also notate local maxima (over a 30-day window). 
 
The primary finding of this figure is a drastic peak on June 6, 2020. This reveals that the spatial distribution of cases throughout the month of May (broadly speaking) was significantly different to that throughout the month of June. Other more moderate peaks are observed on August 23, 2020,  December 6, 2020, March 5, 2021, and May 30, 2021. Throughout the figure, the function (\ref{eq:distancefn1}) displays approximately periodic intervals of increase and decrease. To complement our primary finding, we take a deeper analysis of the 30-day periods prior to and following our peak date, June 6, 2020.

The primary difference between the spatial distribution of COVID-19 cases across the US during the months of May and June 2020 is that May was dominated by high numbers of new cases in the Northeastern states, while cases spread much wider in June, impacting larger states such as California, Texas and Florida with significant case counts. We elucidate and quantify this with an individual state-by-state analysis of the distributional changes between May and June.

\begin{figure*}
    \centering
    \begin{subfigure}[b]{0.32\textwidth}
        \includegraphics[width=\textwidth]{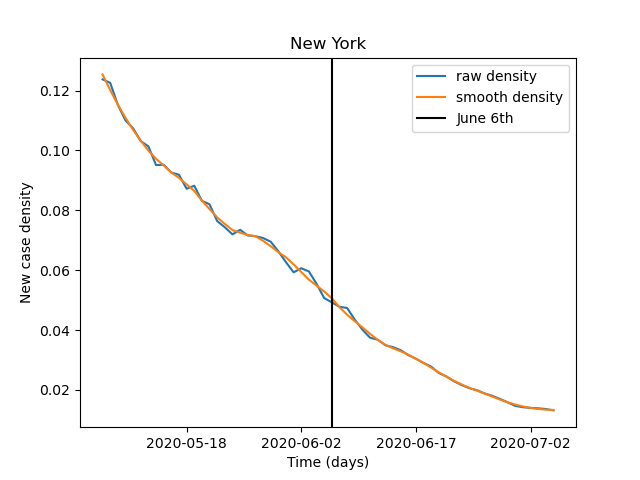}
        \caption{}
        \label{fig:NewYork}
    \end{subfigure}
    \begin{subfigure}[b]{0.32\textwidth}
        \includegraphics[width=\textwidth]{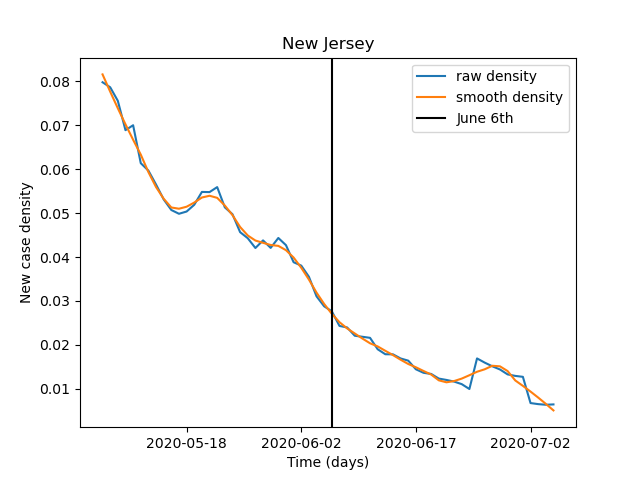}
        \caption{}
        \label{fig:NewJersey}
    \end{subfigure}
    \begin{subfigure}[b]{0.32\textwidth}
        \includegraphics[width=\textwidth]{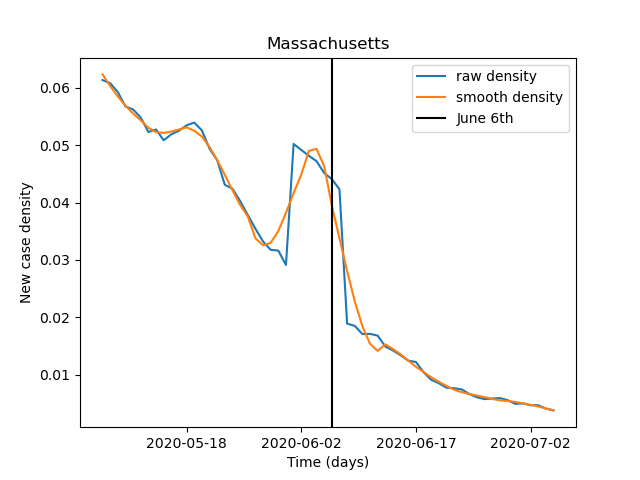}
        \caption{}
        \label{fig:Massachusetts}
    \end{subfigure}
\begin{subfigure}[b]{0.32\textwidth}
        \includegraphics[width=\textwidth]{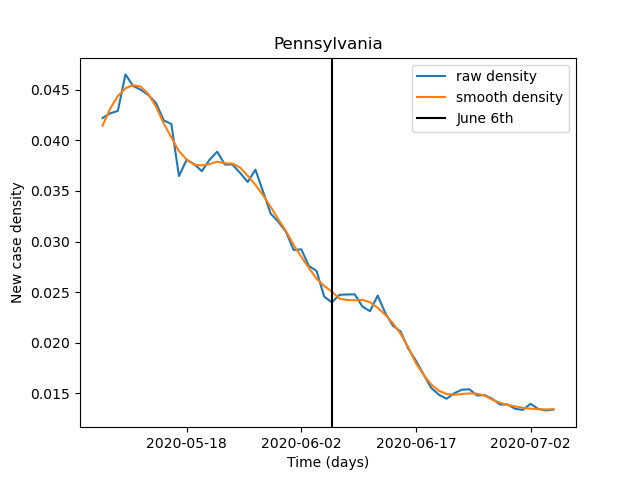}
        \caption{}
        \label{fig:Pennsylvania}
    \end{subfigure}
\begin{subfigure}[b]{0.32\textwidth}
        \includegraphics[width=\textwidth]{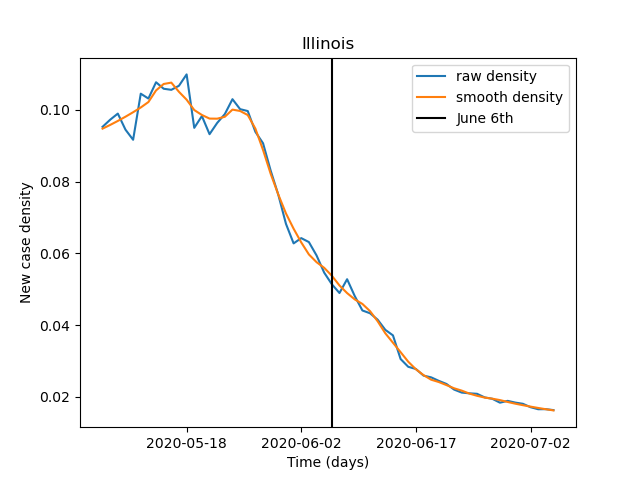}
        \caption{}
        \label{fig:Illinois}
    \end{subfigure}    
        \begin{subfigure}[b]{0.32\textwidth}
        \includegraphics[width=\textwidth]{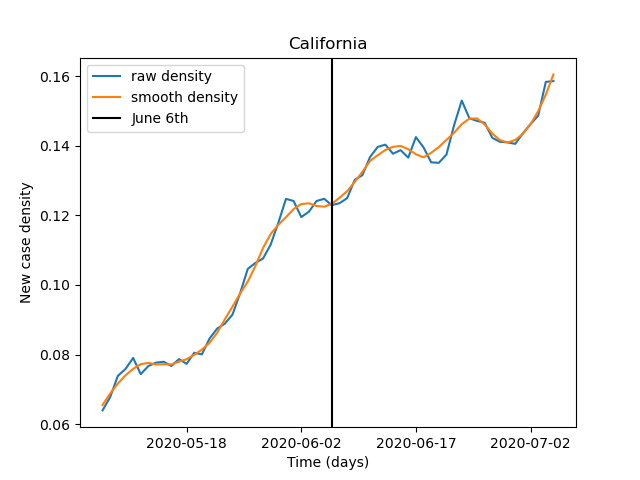}
        \caption{}
        \label{fig:California}
    \end{subfigure}
    \begin{subfigure}[b]{0.32\textwidth}
        \includegraphics[width=\textwidth]{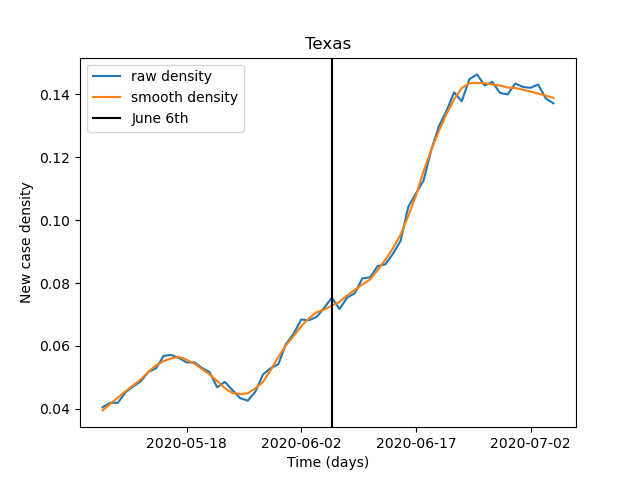}
        \caption{}
        \label{fig:Texas}
    \end{subfigure}
    \begin{subfigure}[b]{0.32\textwidth}
        \includegraphics[width=\textwidth]{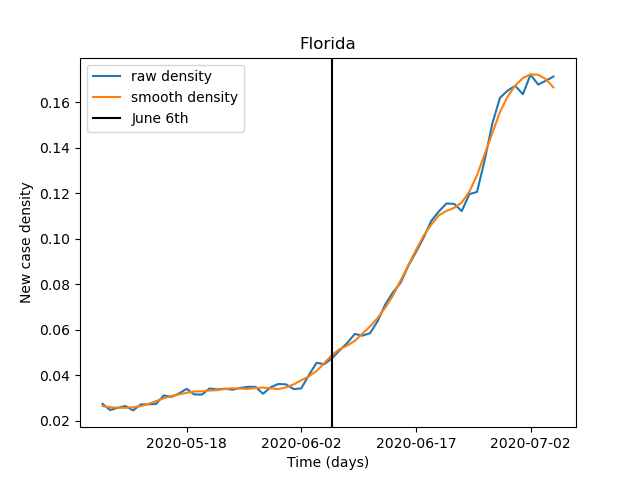}
        \caption{}
        \label{fig:Florida}
    \end{subfigure}
\begin{subfigure}[b]{0.32\textwidth}
        \includegraphics[width=\textwidth]{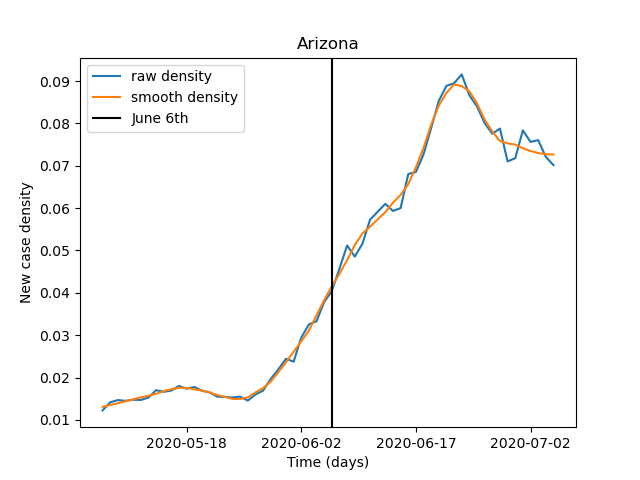}
        \caption{}
        \label{fig:Arizona}
    \end{subfigure}
    \caption{Time-varying densities of new cases for (a) New York (b) New Jersey (c) Massachusetts (d) Pennsylvania (e) Illinois (f) California (g) Texas (h) Florida (i) Arizona across the period May 7 to July 5, 2020. This 60-day interval was identified in fig. \ref{fig:GW_CPD} as the most substantial period of geographic change in the entire period.}
    \label{fig:State_density_time_series}
\end{figure*}

\begin{figure}
    \centering
    \includegraphics[width=0.49\textwidth]{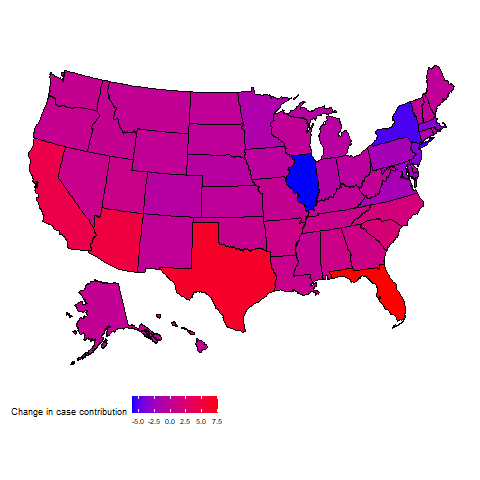}
    \caption{US states and their values of $\Delta(f)_i$, defined in (\ref{eq:delta_i}), measuring the percentage change in their contribution to new US cases between the adjacent 30-day periods of May 7 to June 5 and June 6 to July 5, 2020. California, Texas and Florida exhibit the greatest increase, while New York and Illinois show the greatest decrease.}
    \label{fig:USmap}
\end{figure}

In fig. \ref{fig:State_density_time_series}, we plot the time-varying probability densities $f_i(t)$ for select states over the period May 7, 2020 to July 5, 2020. These figures show the substantial change in the distribution of COVID-19 across the country during this 60-day period. In figs. \ref{fig:NewYork}, \ref{fig:NewJersey}, \ref{fig:Massachusetts}, \ref{fig:Pennsylvania} respectively, we see that four Northeastern states, New York, New Jersey, Massachusetts, and Pennsylvania exhibit a considerable decline in this period in their proportion of the total new COVID-19 cases across the US. While not a Northeastern state, the similarly urbanised and politically inclined state of Illinois (\ref{fig:Illinois}) also exhibits a drastic decline. On the other hand, California (\ref{fig:California}), Texas (\ref{fig:Texas}), Florida (\ref{fig:Florida}) and Arizona (\ref{fig:Arizona}) exhibit a considerable increase. Altogether, fig. \ref{fig:State_density_time_series} shows the decline in the densities in the Northeastern states and the relative increase in all other areas of the country in the month of June.

We provide more quantitative detail in Table 1 of the appendix, where we compute normalised integrals of the time-varying daily density function $f_i(t)$ for each state between May 7 to June 5 and June 6 to July 5. That is, we compute the following for each state, expressed as a percentage: 
\begin{align}
\label{eq:delta_i}
\Delta(f)_i = \left(\frac{1}{30} \sum_{t=t_0}^{t_0 +29} f_i(t) - \frac{1}{30}\sum_{t=t_0 - 30}^{t_0 - 1} f_i(t)\right) \times 100.
\end{align}
The states with the largest absolute changes in their average percentage of US total new cases are Florida (increase of 7.59), Texas (increase of 6.25), Illinois (decrease of 6.08), New York (decrease of 5.42), Arizona (increase of 5.00), California (increase of 4.57), New Jersey and Massachusetts (decreases of 3.53 and 3.48 respectively). We include an interpretable map of all 50 US states and DC in Figure \ref{fig:USmap}, where colours show the magnitude of the $\Delta(f)_i$ quantities. Across all states, the sum of the absolute value of the changes in average proportion is 67.86\%. As this coincides with an $L^1$ norm between distributions, we deduce that $W_{disc}(f^{[t_0 - 30: t_0 -1]},f^{[t_0: t_0 +29]}) = 0.3393$, by (\ref{eq:discreteWass}). This is a substantial difference (its maximal possible difference is 1), and validates the sizeable change in the distribution of cases between these two 30-day periods.

We complete this first analysis with a novel mathematical approach to quantify the spread of a probability distribution across a metric space, while taking the spatial structure into account. Given a distribution $f$ corresponding to a measure $\mu$ on a finite metric space $(X,d)$, let
\begin{align}
\label{eq:geodesicvariance}
\text{Var}(f) &= \int_{X \times X} d(x,y)^2 d\mu(x) d\mu(y) \\ &= \sum_{x, y \in X} d(x,y)^2 f(x)f(y),    
\end{align}
where the second equality is valid when the metric space is discrete, as in this example. We term (\ref{eq:geodesicvariance}) the \emph{geodesic variance} of the distribution $f$. We explain how this is a generalisation of the classic notion of variance in the appendix (Proposition 2).

\begin{figure}
    \centering
    \includegraphics[width=0.49\textwidth]{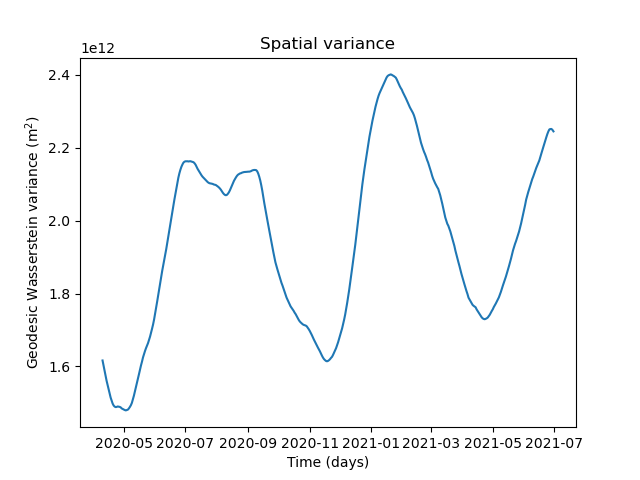}
    \caption{Time-varying geodesic variance (in square meters) (\ref{eq:geodesicvariance}) of 30-day aggregated distributions. The global minimum at May 3, 2020 reflects the spatial concentration of new COVID-19 cases in the Northeastern states during the period May 3 to June 1. In the proceeding 30 days, the geodesic variance grows considerably as new cases become much more spread across the US.}
    \label{fig:GW_Var30}
\end{figure}

The geodesic variance is zero for a Dirac delta $f=\delta_x$, and greater when the distribution is more spread out across the space. In fig. \ref{fig:GW_Var30}, we plot the time-varying variance of grouped 30-day distributions, $t \mapsto \text{Var}(f^{[t: t+29]})$. This figure reveals that the geographic variance of COVID-19 across the US is globally minimal when $t$ corresponds to May 3, 2020, reflecting the 30-day period until June 1, 2020. From this point, it sharply increases, corresponding to the spread of the virus across the country, complementing the findings already observed.

\section{Reversion of spatial propagation over time}

\begin{figure}
    \centering
    \begin{subfigure}[b]{0.49\textwidth}
        \includegraphics[width=\textwidth]{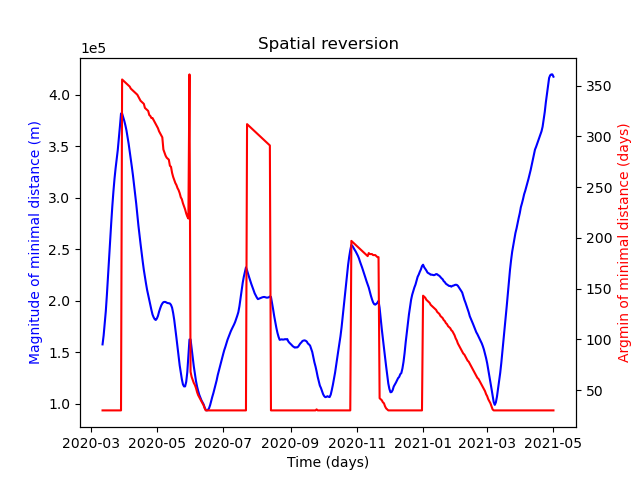}
        \caption{}
        \label{fig:GW_Spatial_evolution}
    \end{subfigure}
    \begin{subfigure}[b]{0.49\textwidth}
        \includegraphics[width=\textwidth]{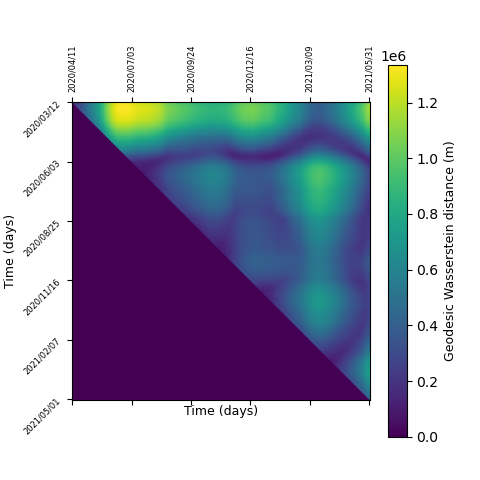}
        \caption{}
        \label{fig:GW_min_distance}
    \end{subfigure}
    \caption{Spatial reversion of the distribution of COVID-19 cases over time. In (a), we plot both the argmin, in days, and the size of the minimal value, in meters, of $\{W_G(f^{[s:s+29]}, f^{[t:t+29]}): s \geq t+30 \}$ as $t$ varies. Periods where the argmin is greater than 30 indicates non-trivial spatial reversion occurs at a time later than $t$. April 2020 is revealed to be weakly similar to a year in the future, but not to any other time. June 2020 is revealed to be similar to both January 2021 and June 2021. In (b), we plot $W_G(f^{[s:s+29]}, f^{[t:t+29]})$ for all non-trivial values $s \geq t+30$. Substantial similarities between June 2020 and various later times are visible in the darker band to the top right of the figure.}
    \label{fig:SpatialEvolutionDensity}
\end{figure}

For our second aim, we investigate the extent of spatial reversion over time. We wish to identify, for some values of $t$, if there is  a non-trivial similarity with subsequent times $s>t$. For each $t$, we consider intervals $\{s: s \geq t+30 \}$ to avoid trivial similarity with neighbouring times and record both $\min \{W_G(f^{[s:s+29]},f^{[t:t+29]}): s \geq t+30 \}$ and $\argmin \{W_G(f^{[s:s+29]},f^{[t:t+29]}): s \geq t+30 \}$. Ignoring the trivial region of being within 30 days of $t$, these determine the time $s$ whose COVID-19 distribution is most similar to time $t$ as well as the magnitude of that similarity. We plot both the argmin function, in days, and the min function, in meters, in fig. \ref{fig:GW_Spatial_evolution}. To provide more details, we also plot $W_G(f^{[s:s+29]},f^{[t:t+29]})$ for all non-trivial values of $s \geq t+30$ in fig. \ref{fig:GW_min_distance}.

In fig. \ref{fig:GW_Spatial_evolution}, non-trivial spatial reversion is characterised by the argmin function being greater than 30 (the nearest permissible day under consideration). With this in mind, two main findings are observed. First, April 2020 has both a high min and argmin, seen in fig. \ref{fig:GW_Spatial_evolution}. The value of the argmin is 350 days, revealing maximal proximity to a period about one year later; the value of the min is approximately 400 km. Thus, this period is not very similar to any other period, but it is weakly similar to April 2021. April 2020 is characterised by the early dominance of the Northeastern US states. This reflects that this period was quite distinct from any other time, but that April 2021 reflects a weak reemergence of the Northeast as significant contributor to the nation's cases.

Secondly, June 2020 is characterised by a relatively low value in the minimum, revealing substantial non-trivial similarity to a subsequent period. The argmin here is approximately 200 days and then immediately after about 360 days. This reflects that June 2020 is highly similar to the period comprising December 2020 and January 2021, as well as the period of June 2021. Indeed, June 2020 is characterised by the second wave of COVID-19 in the US overall, and the first wave in large states California, Texas and Florida, while the beginning of 2021 is characterised by the third wave of COVID-19 in the US overall, and the second wave in the aforementioned large states. The sudden increase in the argmin function around this date reveals additional similarity with June 2021, in which a fourth wave of COVID-19 in the US is beginning. That is, the argmin values around June 2020 reveal two different subsequent reversions in the geographic distribution of COVID-19. This strong similarity is also visible in the darker regions of fig. \ref{fig:GW_min_distance}, to the right of the $y$-value of $t$ as June 1, 2020.

\section{Conclusion}

This paper has introduced a new methodological framework to analyse the spatio-temporal propagation of a process, such as an epidemic, across a physical region. Applying this to the spread of COVID-19 across the US, we have identified periods of rapid change and spatial reversion regarding the country's distribution of COVID-19. In May 2020, new COVID-19 cases were mostly concentrated in the Northeastern states; in June 2020, new cases were spread across the country. During June 2020, the majority of the US was experiencing its second wave of COVID-19; however, California, Texas and Florida, the three largest US states and all outside the Northeast, were experiencing their individual first waves \cite{james2020covidusa}.

Our spatial reversion analysis can capture the broad similarity between June 2020, the second wave of the country as a whole and the first in the aforementioned three states, and January 2021, the third wave in the country as a whole and the second in the three states. A weak similarity is observed between April 2020 and April 2021, signifying a return of the prominence of the Northeastern states.

Future work could modify our framework, replacing geographical distance with other notions of affinity. For example, one could consider two locations as ``close'', and assign them a high similarity, if there is a substantial volume of people travelling between them. Our methodology could be combined with existing methods of network analysis \cite{Hncean2020,Li2021_Matjaz}, which have also been applied to particular regions \cite{Hncean2020_Romania}, or applied on a more granular basis to individual US counties.

Whether combined with other mathematical approaches or used in isolation, our methodology may provide analysts and policymakers alike with additional insights to protect their localities in advance. During the first wave of COVID-19 in the US, the Northeastern states were severely impacted; during this time, several Southern states did not take precautions, with governors drawing considerable differences between their state and New York \cite{AlabamanotNY}. However, our analysis reveals that the geographic distribution of COVID-19 was changing rapidly by the day, and that other locations could have prepared more thoroughly.

Combating COVID-19 requires a collaborative effort between regions that share geographic borders, such as US states and European countries. Studying the time-varying distribution of new COVID-19 cases combined with existing network analyses could provide opportunities for such neighbouring states to work together  combating the ongoing pandemic \cite{Momtazmanesh2020,Priesemann2021}  and new variants \cite{Priesemann2021_actionplan}, and incorporating the strategic use of vaccinations \cite{Markovi2021,Piraveenan2021}.

\section{Data availability statement}

COVID-19 data is sourced from the New York Times \cite{datasourcenyt} and US location data is sourced from Google \cite{datasourceUSgeography}.

\begin{acknowledgments}
The authors would like to acknowledge the reviewers for helpful comments and insights.
\end{acknowledgments}

\clearpage
\onecolumn

\section{Appendix}

\begin{prop}
Let $(X,d)$ be a finite discrete metric space, with $d(x,y)=1$ for all $x\neq y$ and zero otherwise. Let $W^1(\mu,\nu)$ be the $L^1$-Wasserstein metric between two probability measures $\mu,\nu$ on $X$, with corresponding distribution functions $f,g$, expressed as in (2) of the manuscript. That is,
\begin{align}
\label{eq:Wassersteinalt_new}
    W^1 (\mu,\nu) = \sup_{F} \left| \int_{X} F d\mu - \int_{X} F d\nu  \right|.
\end{align}
Then, this supremum is optimized by the choice of $F$ as in (3) of the manuscript, namely
\begin{align}
\label{eq:bestF_new}
F(x) = \begin{cases}
1, & f(x) \geq g(x), \\
0, & f(x)<g(x).
\end{cases}
\end{align}
As such, $W^1(f,g)$ reduces to the simple form of (4) of the manuscript, namely
\begin{align}
\label{eq:discreteWass_new}
W^1(f,g)=\frac12 ||f - g||_1.
\end{align}
\end{prop}
\begin{proof}
Let $F$ be an arbitrary $1$-Lipschitz function on $X$. That is, $F: X \to \mathbb{R}$ and $|F(x) - F(y)| \leq d(x,y) =1$ for $x\neq y$. Let $M=\sup_{x\in X} F(x), m= \inf_{y \in X} F(y).$ By taking the supremum over $x$ and the infimum over $y$, the Lipschitz condition ensures that $M-m\leq 1.$ So
\begin{align}
  \int_{X} F d\mu - \int_{X} F d\nu  &=\sum_{x \in X} F(x)(f(x)-g(x)) \\
  &\leq \sum_{x: f(x) \geq g(x) } M(f(x)-g(x)) + \sum_{x: f(x) < g(x) } m(f(x)-g(x))\\
  &\leq \sum_{x: f(x) \geq g(x) } (m+1)(f(x)-g(x)) + \sum_{x: f(x) < g(x) } m(f(x)-g(x))\\
  &=\sum_{x: f(x) \geq g(x) } f(x)-g(x) + \sum_{x \in X}m(f(x)-g(x))\\
  &=\sum_{x: f(x) \geq g(x) } f(x)-g(x),
\end{align}
using the fact that $\sum_x f(x)=\sum_x g(x)=1$ to eliminate the second sum. Now, let
\begin{align}
    P&=\sum_{x: f(x) \geq g(x) } f(x)-g(x),\\
    N&=\sum_{x: f(x) < g(x) } g(x)-f(x).
\end{align}
Then $P-N = \sum_{x\in X} f(x)-g(x) = 0,$ while $P+N = \sum_{x \in X} |f(x)-g(x)|=\|f-g\|_1.$ Thus, we deduce $P=N=\frac12 \|f-g\|_1,$ and $\left|\int_{X} F d\mu - \int_{X} F d\nu\right| \leq P$. Taking the supremum over $F$, we deduce $W^1(f,g) \leq P$. Finally, let $F$ be as in (\ref{eq:bestF_new}). Then $\int_{X} F d\mu - \int_{X} F d\nu$ immediately coincides with $P$, by definition. Thus, the supremal value coincides with $P$, and $P=\frac12 \|f-g\|_1$, as required.

\end{proof}

\begin{prop}
Let $f$ be a probability distribution on a finite metric space $(X,d)$, and consider its geodesic variance, defined by (8) of the manuscript. That is,
\begin{align}
\label{eq:geodesicvariance_new}
\text{Var}(f) &= \int_{X \times X} d(x,y)^2 d\mu(x) d\mu(y)\\ &= \sum_{x, y \in X} d(x,y)^2 f(x)f(y).  
\end{align}
When $(X,d)$ is a finite subset of the real numbers $\mathbb{R}$ with its Euclidean metric, this quantity reduces to the classical notion of variance, up to a factor of 2.
\end{prop}
\begin{proof}
Let $Y$ be a random variable on $\mathbb{R}$. The classical notion of variance is the quantity $\text{var}(Y)=\E Y^2 - (\E Y)^2$. If $Y$ has a distribution $f$ over a finite set $X$, this can be expressed
\begin{align}
    \text{var}(Y) = \sum_{x \in X} x^2 f(x) - \left(\sum_{x \in X} x f(x)\right)^2.
\end{align}
On the other hand, (\ref{eq:geodesicvariance_new}) can be expanded
\begin{align}
    \text{Var}(f) &= \sum_{x, y \in X} (x-y)^2 f(x)f(y)\\
   &=\sum_{x, y \in X} x^2 f(x)f(y) + \sum_{x, y \in X} y^2 f(x)f(y) - 2\sum_{x, y \in X} xyf(x)f(y)\\
     &= 2\sum_{x, y \in X} x^2 f(x)f(y) - 2\sum_{x, y \in X} xyf(x)f(y)\\
    &=2\sum_{x \in X} x^2 f(x) - 2 \left(\sum_{x \in X} x f(x)\right)\left(\sum_{y \in X} y f(y)\right)\\
    &=2\text{var}(Y).
\end{align}
Thus, up to a factor of 2, the geodesic variance reduces to the classical notion of variance on the real line. 
\end{proof}

With more care, the above propositions hold if $X$ is an arbitrary metric space, with $\mu, \nu$ appropriately integrable measures on $X$.

\newpage

In Table \ref{tab:State_density_deltas}, we record all values of $\Delta(f)_i$, as defined in eq. (7) and depicted in fig. 3 of the manuscript.

\begin{table}
\begin{center}
\begin{tabular}{ |p{2.2cm}|p{1.2cm}||p{2.5cm}|p{1.0cm}|}
 \hline
 State & $\Delta(f)_i $ & State & $\Delta(f)_i $ \\
 \hline
 Alabama & 0.82 & Montana & 0.04 \\
 Alaska & 0.06 & Nebraska & -0.70 \\
 Arizona & 5.00 & Nevada & 0.62 \\
 Arkansas & 0.93 & New Hampshire & -0.19 \\
 California & 4.57 & New Jersey & -3.53 \\
 Colorado & -0.75 & New Mexico & -0.09 \\
 Connecticut & -1.41 & New York & -5.42 \\
 Delaware & -0.41 & North Carolina & 1.58 \\
 D.C. & -0.38 & North Dakota & -0.10 \\
 Florida & 7.59 & Ohio & -0.51 \\
 Georgia & 1.09 & Oklahoma & 0.42 \\
 Hawaii & 0.03 & Oregon & 0.30 \\
 Idaho & 0.22 & Pennsylvania & -1.89 \\
 Illinois & -6.08 & Rhode Island & -0.58 \\
 Indiana & -0.94 & South Carolina & 1.96 \\
 Iowa & -0.46 & South Dakota & -0.15 \\
 Kansas & -0.17 & Tennessee & 0.62 \\
 Kentucky & -0.04 & Texas & 6.25 \\
 Louisiana & 0.59 & Utah & 0.70 \\
 Maine & -0.07 & Vermont & 0.01 \\
 Maryland & -2.53 & Virginia & -1.91 \\
 Massachusetts & -3.48 & Washington & 0.23 \\
 Michigan & -0.65 & West Virginia & -0.02 \\
 Minnesota & -1.36 & Wisconsin & -0.09 \\
 Mississippi & 0.13 & Wyoming & 0.02 \\
 Missouri & 0.24 & & \\
\hline
\end{tabular}
\caption{US states and their values of $\Delta(f)_i$, defined in (7) of the manuscript, measuring the percentage change in their contribution to new US cases between the adjacent 30-day periods of May 7 to June 5 and June 6 to July 5, 2020.}
\label{tab:State_density_deltas}
\end{center}
\end{table}

\bibliography{_references}
\bibliographystyle{eplbib.bst}

\end{document}